\documentstyle [11pt]{report}
\def\ZzZ{{\hbox{\tenrm Z\kern-.31em{Z}}}}
\def\CcC{{\hbox{\tenrm C\kern-.45em{\vrule height.67em width0.08em depth-
.04em
\hskip.45em }}}}

\newcommand{\ep}{\epsilon}

\newcommand{\lab}{\label}

\newcommand{\bc}{\begin{center}}
\newcommand{\ec}{\end{center}}
\newcommand{\be}{\begin{equation}}
\newcommand{\ee}{\end{equation}}
\newcommand{\bea}{\begin{eqnarray}}
\newcommand{\eea}{\end{eqnarray}}
\newcommand{\bs}{\begin{subequations}}
\newcommand{\es}{\end{subequations}}
\newcommand{\beq}{\begin{eqalignno}}
\newcommand{\eeq}{\end{eqalignno}}

\def\vec#1{\mbox{\boldmath $#1$\unboldmath}}

\def\vec#1{{\bf #1}}

\def\lab{\label}
\def\lan{\langle}

\def\le{\left}

\def\Lrar{\Leftrightarrow}

\def\pa{\partial}
\def\ran{\rangle}
\def\rar{\rightarrow}

\def\ri{\right}
\def\ti{\tilde}

\def\de{\delta}
\def\De{\Delta}
\def\ep{\epsilon}

\def\te{\theta}

\def\la{\lambda}

\def\lab{\label}

\begin{document}

$$ $$

\bc
{
{\bf ON TOPOLOGICAL DEFECT FORMATION IN THE PROCESS OF
SYMMETRY BREAKING PHASE TRANSITIONS}

$$ $$


Eleonora Alfinito, Oreste Romei and Giuseppe Vitiello

\bigskip


Dipartimento di Fisica,
INFN and INFM

Universit\`a di Salerno, 84100, Italia

e-mail: alfinito@sa.infn.it, romei@sa.infn.it, vitiello@sa.infn.it

$$ $$
}

\ec

{\bf Abstract} By resorting to some results in quantum field theories with spontaneous breakdown of symmetry we show that an explanation
based on microscopic dynamics can be given of the fact that topological defect formation is observed during the process of
non-equilibrium phase transitions characterized by a non-zero order parameter. We show that the Nambu-Goldstone particle acquires an effective non-zero mass due to boundary (finite volume) effects and this is related with the size of the defect. We also relate such volume effects with temperature effects.

$$  $$


$$   $$

Much attention is currently devoted to the problem of topological defect
formation during the process of symmetry breaking phase
transitions \cite{Bunkov}. In such a process it may happen that a region, surrounded by ordered domains, remains trapped in the "normal" or symmetric state. This occurrence manifests as a topological defect. Topological defects are observed as macroscopic
extended objects with classical behavior, e.g. vortices in
superconductors and superfluids, magnetic domain walls in ferromagnets,
dislocations, grain boundaries, point defects in crystals. In cosmology,
topological defects, such as cosmic strings, may have been playing a r\^ole
in the phase transition processes in the early Universe \cite{kib}. The Kibble-Zurek scenario \cite{kib2, zurek1} provides the phenomenological understanding of the defect formation in phase transitions.
There is
a surprising analogy between defect formation in solid state physics and in
high energy physics and cosmology \cite{volovik1}. For an interesting table
of analogies see \cite{volovik2}.
As an example, we just mention the analogy between vortex
in superfluids and global cosmic strings. Studying the physics of defect
formation in condensed matter physics may be then helpful in the
understanding of possible scenarios in the early Universe
cosmology \cite{Bunkov}. The analysis of the formation of defects in phase
transitions thus becomes a "diagnostic tool" \cite{zurek} in the study of
non-equilibrium symmetry breaking processes in a wide range of energy
scale.

In this paper, by limiting ourselves to the class of non-equilibrium symmetry breaking phase transitions characterized by an order parameter,
we discuss some of the questions which arise
in  the study of defect formation. For example, which one is the
connection  between defect formation and symmetry breaking, i.e., more
precisely, why  are extended objects with topological singularity
observed only in systems showing some  sort of ordered patterns? Why
defect formation is observed during the  processes of phase
transitions? Why are the features of the defect formation so general,
i.e. why are they shared by quite different systems, from condensed
matter to cosmology? How to explain in a  unified theoretical scheme
the macroscopic behavior of extended objects and  their interaction
with quanta? How temperature and finite volume (the  system
boundaries) effects fit in such a theoretical frame?

Of course,  there
is a large body of literature in quantum field theory (QFT) on the
dynamics of defect formation and some of the above questions have been, at least partially, answered. By resorting indeed to some of such works,
in the present paper we reach the following main results:

i) we show that a theoretical explanation
based on microscopic dynamics can be given of the fact that topological defect formation is observed during the process of non-equilibrium symmetry breaking phase transitions (limited to the case of the existence of the order parameter and in the presence of a gauge field).

ii) we provide an explicit proof that the Nambu-Goldstone (NG) particle acquires an effective non-zero mass due to boundary (finite volume) effects, which is related with the size of the defect.
We also relate such volume effects with temperature effects.

Our derivation is based on symmetry properties
of the dynamics at the basic level of the quantum fields.
As far as we know, in the literature the conclusions we reach are
commonly accepted mostly on the basis of phenomenological arguments. Also, our conclusions are model independent, which may explain why some of the features of defect formation, within the limitations imposed to our analysis, are shared by many systems in a wide range of energy scale,
independently of specific aspects of the system dynamics.
Moreover, our results apply to the case of the Abelian U(1) symmetry (for vortices) as well as to the non-Abelian case of SO(3) and SU(2) symmetry (as for the monopole and the sphaleron case, respectively). As already mentioned, throughout this paper our discussion is limited to the case in which an order parameter exists and a gauge field is present. However, we also present some comments on boundary and temperature effects on the kink defect.

We arrive at the result i) as a direct consequence of two known properties of gauge theories with spontaneous breakdown of symmetry (SBS). Both these properties have to do with the topological characterization of the
non-homogeneous boson condensation in SBS theories. Although these properties are known since several years, nevertheless they have not been considered till now in relation with the problem of the appearance of topological defects in phase transitions. For a preliminary discussion see ref. \cite{les}. For the reader convenience we report in the Appendices some details of the mathematical formalism to which we have to resort in our discussion.

Let us start by recalling that in QFT the formation of extended objects
is described by
non-homogeneous boson condensation \cite{Um2} - \cite{MPU75}. This condensation is formally obtained by translations
of boson fields (not necessarily massless), say
$\chi_{in}(x) \rar \chi_{in} (x) + f(x)$, with c-number function $f(x)$,
{\it satisfying the
same field equation for $\chi_{in} (x)$}. These translations are called
{\it boson transformations} \cite{Um2}. The
boson transformation function plays the r\^ole of a "form factor" and
the extended object appears as the macroscopic envelop of the
non-homogeneous boson condensate (localized over a finite domain). The
topological charge of the extended object (the defect) arises from the
topological singularity of the boson condensation  function.

We remark that, on the other hand, transitions between
the system phases characterized by different ordered patterns
in the ground
state are induced by variations (gradients) of the NG boson
condensate, i.e. of boson transformation function. Hence, in order to show that in such a process the conditions are met for the formation of topological defects, we need

(a) to consider under which constraints the boson transformation function $f(x)$ can carry a topological singularity and

(b) to show that in the process of phase transitions under study these constraints are in fact satisfied.

Let us first consider the point (a). We recall that in theories where boson condensation can occur, {\it topological defects are observed only in systems exhibiting ordered patterns} (characterized by a non-zero order parameter).

In fact one can show that boson transformation functions carrying
topological singularities are allowed only for massless bosons \cite{Um2, Um1}, such as NG bosons of SBS theories where ordered ground states appear.   The proof of this goes as follows \cite{Um1}.

Consider the boson transformation $\chi_{in}(x)  \rar  \chi_{in}(x) +
f(x)$. Let $f(x)$ carry a topological singularity, which means that
it is path-dependent:
\be
\\ \lab{ts1}
G^{\dag}_{\mu\nu}(x) \equiv [\pa_\mu,\pa_\nu]\,f(x) \neq 0~,
\qquad for \;certain \; \mu\, , \,  \nu
\, , \, x  ~.
\ee

We will see below that $\pa_\mu \, f$ is related with
observables and therefore it is single-valued,
i.e. $[\pa_\rho,\pa_\nu]\,\pa_\mu f(x)\,=\,0$. Recall that
$f(x)$ is solution of the $\chi_{in}$ equation and suppose there is
a non-zero mass term: $(\pa^2 + m^2)f(x) = 0$. From the regularity of
$\pa_\mu f(x)$ it follows that
\be\lab{ts5}
\pa_\mu f(x) \, =\, \frac{1}{\pa^2 \, + \, m^2}
\pa^{\la} \,G^{\dag}_{\la\mu}(x) ~,
\ee
which leads to
$\pa^2 f(x) \, =\,0$, which in turn implies $m=0$.
{\it Thus (\ref{ts1}) is  compatible only with
massless $\chi_{in}$}. This explains why topological defects are
observed only in systems exhibiting ordered patterns,
namely in the presence of NG boson condensation sustaining the long range ordering correlation.

We remark that the above conclusion is not limited by dimensional considerations or by the Abelian or the non-Abelian nature of the theory symmetry group. It can be easily extended to the non-relativistic regime (see \cite{Um1}) and it holds true in a theory with or without gauge fields.  It applies to a full set of topologically non-trivial extended objects, ranging from topological line singularity, to surface singularity, point defects, grain boundaries and dislocation defects in crystals, SU(2)-triplet model and monopole singularity. The topological singularity
and the topological charge of the related extended object can be completely characterized.For a detailed account see ref.
\cite{Um1}, especially Chapter 10.

Next, for the point (b), we consider the case in which a gauge field is present. We then observe that from Eq. (\ref{Bvs20}) (Appendix  B) we obtain the classical ground state current $j_{\mu}$ as
\be\lab{vs21}
j_\mu(x)\equiv \lan 0| j_{H \mu}(x) |0 \ran \, =\,
 m_V^2 \le[ a_\mu(x) - \frac{1}{e_0} \pa_\mu f(x) \ri]~,
\ee
where $ m_V^2  a_\mu(x)$ is the {\em Meissner-like current} and
$ \frac{m_V^2}{e_0} \pa_\mu f(x)$ is the {\em boson current}.

Note that the observable classical current is
given in terms of gradients ${\pa}_{\mu}f$ of the
boson condensation function. Eq. (\ref{vs21}) shows that the macroscopic ground state effects, such as the classical field and the classical current, do not occur for regular $f(x)$ ($G^{\dag}_{\mu\nu} = 0$). In
fact, from (\ref{Bvs20}) we obtain
${\pa^2} a_\mu(x) = 0$ for regular $f(x)$, i.e.
$a_{\mu}(x) = \frac{1}{e_{0}}
\pa_{\mu} f(x)$, which in turn implies zero classical
current ($j_{\mu} = 0$) and zero classical field
($F_{\mu\nu} = \pa_{\mu} a_{\nu} -  \pa_{\nu} a_{\mu}$) (the
Meissner-like current and the boson current cancel each other).

The conclusion is that the vacuum current appears only when $f(x)$ has
topological singularities, which, as we have seen above (point (a)), is only compatible with the condensation of massless bosons,
i.e. when SBS occurs. Again this conclusion is a general one, it does not depend on the specific model. It holds for Abelian as well as for non-Abelian groups, in the relativistic and in the non-relativistic case.

{\it Summing up, in a gauge theory,
the symmetry breaking phase
transitions characterized by macroscopic ground state effects, such
as the vacuum current and field, (e.g. in superconductors)
can occur only  when there are non-zero gradients of
topologically non trivial condensation of NG bosons}.
But this means that the conditions for the
formation of topological defects are met and {\it thus we see why
topological defects are observed in the process of symmetry breaking
phase transitions}. We have thus obtained our result i).

Notice that the appearance of space-time dependent order parameter
$\ti v$ is no guarantee that persistent ground state currents (and
fields) will exist: if $f$ is a regular function, the space-time
dependence of $\ti v$ can be gauged away by an appropriate gauge
transformation. Which means that non-homogeneous condensation, and thus  extended objects, cannot be obtained in such case. Also note that in a
theory which has only global gauge invariance non-singular boson
transformations of the NG fields can produce
non-trivial physical effects (like linear flow in superfluidity).

In conclusion, in a gauge theory with non-zero order parameter, in order to have topological defects one needs boson condensation function with topological singularities and this is only allowed for NG boson condensation. These conditions are all met in the process of phase transition in such a theory.

A further consequence of the above analysis is that,
since the boson transformation with
regular $f$ does not affect observable quantities, the $S$
matrix must be actually given by (cf. Appendix A)
\be\lab{lp56aa}
S\,=\,: S[\rho_{in}, U^\mu_{in} - \frac{1}{m_V} \pa(\chi_{in} -
 b_{in})] : ~.
\ee
This is in fact independent of the boson transformation with
regular $f$:
\be\lab{lp56ab}
S\,\rar\,S' = : S[\rho_{in}, U^\mu_{in} - \frac{1}{m_V}
\pa(\chi_{in} - b_{in})
+ Z^{-\frac{1}{2}}_{3} (a^{\mu} - \frac{1}{e_{0}} {\pa}^{\mu} f)]:
\ee
since  $a_{\mu}(x) = \frac{1}{e_{0}}
\pa_{\mu} f(x)$ for regular $f$. However, $S'
\neq S$ for singular $f$: in such a case (\ref{lp56ab}) shows that
$S'$ includes the
interaction of the quanta
$U^\mu_{in}$ and $\rho_{in}$ with the classical field
and current.
This shows how it may happens that {\it quanta interact and have effects on
classically behaving macroscopic extended objects}.

For the point ii), let us consider now the effects of finite volume on homogeneous boson condensation.

For large but finite volume we expect that the
symmetry breakdown order parameter is constant
``inside the bulk'' {\em far} from the
boundaries. However, ``{\em near}'' the boundaries, one might
expect ``distortions'' in the order parameter: ``near''
the system boundaries we may have non-homogeneous
order parameter, ${\ti v} ={\ti
v}(x)$ (or even ${\ti v}\rar 0$). Such non-homogeneities in the boson condensation
will ``smooth out'' in the $V\rar\infty$ limit.

{}From the Ward-Takahashi identities one obtains the following pole
structure for the
two-point function of the $\chi(x)$ field \cite{MPUV75}:
\be\lab{lp160}
 \lan \chi(x)\chi(y)\ran = \lim_{\ep\rar 0} \le\{
\frac{i}{(2\pi)^4} \int d^4p\,
\frac{Z_{\chi} e^{-i p (x-y)}}{p^2 - {m_{\chi}}^{2}+i \ep a_\chi}
+ (continuum \;contributions) \ri] ~.
\ee
$Z_\chi$ and $a_\chi$ are renormalization constants.
The space integration of $\lan \chi(x)\chi(y)\ran$ then picks up the pole
contribution at
$p^2=0$, and leads to \cite{MPU741, MPU742}
\be\lab{gp9}
{\ti v}= \frac{Z_\chi}{a_\chi} v \Lrar m_\chi = 0 ~,~~~ or ~ ~~
{\ti v}= 0 \Lrar m_\chi \neq 0   ~,
\ee
where $v$ denotes a convenient c-number \cite{MPUV75}.
Eq. (\ref{gp9}) expresses the well known Goldstone theorem:
if the symmetry is
spontaneously broken (${\ti v} \neq 0$),
the NG massless mode exists, whose interpolating Heisenberg field
is $\chi_{H} (x)$. Since it is massless it
manifests as a long range correlation mode and it is thus
responsible for the vacuum ordering, as already observed above.

Restrict now the space integration of Eq. (\ref{lp160}) over the
finite (but large) volume $V \equiv \eta ^{-3}$ and use for each
space component of $p$:
\be\lab{note13}
\de_\eta(p)=\frac{1}{2\pi}\int_{-\frac{1}{\eta}}^{\frac{1}{\eta}}
dx\,e^{ipx}=\frac{1}{\pi p}\, sin\frac{p}{\eta}  ~,
\ee
which, as well known, approaches $\de (p)$ as $\eta\rar 0$: $
lim_{\eta\rar 0} \de_\eta(p)=\de (p) $. Consider that
\be\lab{note15}
lim_{\eta\rar 0} \int dp\, \de_\eta(p)\,f(p)=f(0)=
lim_{\eta\rar 0} \int dp\, \de(p-\eta)\,f(p)  ~.
\ee
Then, using
$\de_\eta(p)\simeq\de(p-\eta)$ for small $\eta$, one obtains
\be\lab{6}
{\ti v}(y,\ep , \eta)=i{\ep} v e^{-i\vec{\eta}\cdot\vec{y}}\,
\Delta_{\chi}(\ep,\vec{\eta},p_0=0)  ~,
\ee
with
\be\lab{gp8}
\De_\chi(\ep, \vec{\eta}, p_{0}=0) =
\le[ \frac{Z_\chi}{{- \omega_{\vec{p} =
\vec{\eta}}^{2}}
+i\ep  a_\chi} + (continuum \;contributions) \ri]\, ,
\ee
and ${\omega^{2}}_{\vec{p} = \vec{\eta}} = \vec{\eta}^{2} +
{m_{\chi}}^{2}$. Thus, ${lim_{\ep \rar 0}} {lim_{\eta \rar
0}} {\ti v}(y, \ep, \eta) \neq 0$ only if $m_{\chi} = 0$,
otherwise ${\ti v} = 0$. Note that the Goldstone theorem
is thus recovered in the infinite volume limit ($\eta \rar 0$).

On the other hand, if $m_{\chi} = 0$ and
$\vec{\eta}$ is given a non-zero value (i.e. by reducing to a finite volume, i.e. in the
presence of boundaries),
then ${\omega}_{\vec{p} = \vec{\eta}} \neq 0$
and it acts as an "effective mass" for the $\chi$ bosons.
Then, in order to have the order
parameter ${\ti v}$ different from zero $\ep$ must be kept non-zero.

In conclusion, near the boundaries ($\eta \neq 0$) the NG bosons acquire an effective mass
$m_{eff} \equiv {\omega}_{\vec{p} = \vec{\eta}}$.
They will then propagate over a range of the order of
$\xi\equiv\frac{1}{\eta}$, which is the linear size of the condensation domain, or, in the presence of topological singularity, the size of the topologically non-trivial condensation, namely of
the extended object (the defect).

We stress that if  $\eta \neq 0$ then $\ep$ must be
non-zero in order to have the order parameter different from
zero: ${\ti v} \neq 0$ (at least locally). In
such a case the symmetry breakdown is maintained because
$\ep \neq 0$: $\ep$ acts as the coupling with an external field (the
pump) providing
energy. Energy supply is required in order to condensate
modes of non-zero lowest energy ${\omega}_{\vec{p} = \vec{\eta}}$.
In summary, boundary effects are in competition  with  the
breakdown of symmetry \cite{les}. They may preclude its
occurrence or, if
symmetry is already broken, they may reduce to zero the order
parameter.

Temperature may have similar effects on the
order parameter (at $T_{C}$ symmetry may be restored).
Since the order
parameter goes to zero when NG modes acquire non-zero effective
mass (unless, as seen, external energy is supplied),
we may then represent the effect of thermalization in terms of
finite volume effects and put, e.g., $\eta \propto
{\sqrt{\frac{|T-T_C|}{T_C}}}$. In this way temperature
fluctuations around $T_{C}$ may produce fluctuations in the
size $\xi$ of the condensed domain (the size of ordered domain and/or
of the domain where non-homogeneous condensation occurs, namely of
the defect).

For example, suppose to start with $T > T_{C}$, but near to
$T_{C}$. In the presence of an
external driving field ($\ep \neq 0$), even {\it before}
transition to fully ordered phase is
achieved (as $T \rar T_{C}$), one may have the formation of
ordered domains of size
$\xi \propto ({\sqrt{\frac{|T-T_{C}|}{T_C}}})^{- 1}$.
As far as $\eta \neq 0$,
the ordered domains (and the topological defects) are unstable,
they disappear as the
external field coupling $\ep \rar 0$.

Of course, assuming $T$ is lowered below $T_C$, if ordered domains are still present at $T<T_C$,
they also disappear as $\ep\rar 0$.
The possibility to maintain such ordered
domains below $T_C$ depends on the speed at which $T$
is lowered, compared to the speed at which the system is able to
get homogeneously ordered. Notice that the speed at which
$T \rar T_C$ is related to the speed at which $\eta \rar 0$.

In the case of the kink solution, the
boson transformation function $f_{\beta}(x)=const.\, e^{-\mu_0(\beta)x_1}$ plays  the role of ``form
factor'' for the soliton solution. The number of condensed bosons is proportional to
$\vert f_{\beta}(x)\vert^2 = e^{-2\mu_0(\beta)\,(x_1-a)}$,
which is maximal near the kink center $x_1=a$ and decreases over a size
$\xi_\beta=\frac{2}{\mu_0(\beta)}$.
The mass
$\mu_0=(2\la )^{\frac{1}{2}}v(\beta)$, with
\cite{MV90} $v^2(\beta)={\ti v}^2-3\lan:\rho^2:\ran_0< {\ti v}^2$,
of the "constituent" fields $\rho^{in}(x)$ fixes the kink
size $\xi_\beta\propto\frac{2}{\mu_0}=\frac{\sqrt{2}}
{\sqrt{\la}v(\beta)}$,
which thus increases as $T$ increases: $T\rar T_C$ (say
$T\not= T_C$ but near $T_C$. At $T_C$, $v(\beta_{C}) = 0$.).

In the $T \rar 0$ limit the kink size is
$\xi_0\propto\frac{\sqrt{2}}{\sqrt{\la}{\ti v}}<
\frac{\sqrt{2}}{\sqrt{\la}v(\beta)}=\xi_\beta$.
For $T$ different from zero, the
thermal Bose condensate $\lan:\rho^2:\ran_0$ develops, which acts as
a potential term for the
kink field.
It is such a potential term which actually controls
the ``size'' (and the number) of the
kinks. Only far from the kink core, namely in the limit $v(x,\beta)\rar const$,
the $\rho^{in}(x)$ field may be considered as a free field. The translation of the
boson field by $f_{\beta}$ breaks
the homogeneity of the otherwise constant in space order
parameter $v(\beta)$.

Let us close by observing that in the case of topologically
non-trivial
condensation at finite temperature the order parameter
$v(x, \beta)$ provides a mapping between the variation domains
of $(x, \beta)$ and the {\it space of the
unitarily inequivalent representations} of the canonical
commutation relations, i.e. the set of Hilbert spaces where the
operator fields are realized for different values of the
order parameter. This expresses the known fact that we have
non-trivial homotopy mappings between the $(x,\beta)$ variability
domain and the group manifold. As well known, in the vortex case
one has the
mapping $\pi$ of $S^{1}$, surrounding the $r = 0$ singularity, to
the group manifold of U(1) which is topologically characterized
by the winding number $n \in Z \in
\pi_{1}(S^{1})$. It is such a singularity which is carried by the
boson condensation function of the NG modes.  In
the monopole
case \cite{MV90}, the mapping $\pi$
is the one of the sphere $S^{2}$, surrounding the singularity
$r=0$, to SO(3)/SO(2) group manifold, with homotopy classes of
$\pi_{2}(S^{2})= Z$. Same situation
occurs in
the sphaleron case \cite{MV90},
provided one replaces SO(3) and SO(2) with
SU(2) and U(1), respectively.

In conclusion, transitions between phases characterized by an order parameter imply "moving"
over unitarily inequivalent representations, and this is
achieved by gradients in NG boson condensation function. In the presence of
a gauge field, macroscopic ground state field and
currents can only be obtained by non-homogeneous NG boson condensation with
topological singularities. In turn, the occurrence of such topologically non-trivial condensation allows the formation of topological defects: this
explains why topological defect formation is observed in such a kind of symmetry breaking
phase transition processes ({\it "where the defects come from"}).

We have also seen that finite volume (the presence of boundaries) is
related to the size of the
ordered domains and of the defects and we have briefly discussed the correlation with temperature effects. This may justify the expectation relating the number of observed defects with the non-equilibrium
dynamics of phase transitions.

In the case of the kink
there are no NG modes, nevertheless the
topologically non-trivial kink solution requires the boson
condensation function to carry divergence
singularity (at spatial infinity).

This work has been partially supported by INFN,
INFM, MURST and the ESF Network on Topological defect formation in
phase transitions.



\newpage

{\it Appendix A. The boson condensation and the dynamical rearrangement of symmetry}

\vspace{0.5cm}

In the standard Lehmann-Symanzik-Zimmermann
(LSZ) formalism \cite{itz} of  QFT, the dynamics is given in terms of the
interacting fields, say $\varphi_{H} (x)$, also called the
Heisenberg fields. On the other hand,
observables are described by asymptotic in- (or out-)
operator fields (physical fields),
say $\varphi_{in} (x)$, which satisfy free field equations.
The mechanisms described in this appendix are fully general, however, for simplicity we consider the Lagrangian density ${\cal
L}[\phi_{H}(x), {\phi_{H}}^*(x), A_{H \mu}(x)]$
for a complex scalar field $\phi_{H}(x)$
interacting with a gauge field $A_{H \mu}(x)$.
We do not need to specify the detailed structure of the
Lagrangian. We only
require that such a Lagrangian be invariant under global
and local U(1) gauge transformations (as, e.g., in the Higgs-Kibble
model \cite{hig, ki}):
\be\lab{lp1}
\phi_{H}(x) \rar e^{i \te} \phi_{H} (x) \qquad, \qquad \qquad
A_{H \mu}(x) \rar A_{H \mu}(x)~,
\ee
\be\lab{lp2}
\phi_{H} (x) \rar e^{i e_0 \la(x)} \phi_{H} (x) \qquad, \qquad \qquad
A_{H \mu} (x) \rar A_{H \mu} (x) \, + \, \pa_\mu \la(x)~,
\ee
respectively, with $\la(x)\rar 0$ for $|x_0|\rar \infty$ and/or
$|{\bf x}|\rar \infty$. We use the Lorentz gauge
$\pa^\mu A_{H \mu} (x)\,=\,0$ and put
$\phi_{H} (x)=\frac{1}{\sqrt{2}}\le[\psi_{H} (x) + i \chi_{H} (x)\ri]$.
We also assume that SBS can occur:
$\lan 0| \phi_H(x)|0\ran \equiv {\ti v} \neq 0$,
with ${\ti v}$ constant and put $\rho_{H} (x) \equiv \psi_{H} (x) - {\ti v}$.
It is known \cite{MPUV75} that the theory contains a
massless negative norm field (ghost) $b_{in} (x)$, the
Nambu-Goldstone massless mode $\chi_{in} (x)$, and a massive vector field
$U^\mu_{in}$. The relation between the dynamics and the
observable properties of the physical states is provided by the LSZ mapping
(the dynamical map, or Haag expansion) between
interacting fields and physical
fields. In the present case one finds \cite{MPUV75} the following LSZ maps:
\be\lab{lp57a}
\phi_H(x)= :\exp\le\{i  \frac{Z_\chi^{\frac{1}{2}}}
{ {\ti v}}\chi_{in}(x) \ri\}
\le[{\ti v} + Z_\rho^{\frac{1}{2}} \rho_{in}(x) + F[\rho_{in},
U^\mu_{in}, \pa(\chi_{in} - b_{in})] \ri]:
\ee
\be\lab{lp57b}
A^{\mu}_{H}(x)={Z_3^{\frac{1}{2}}} U^{\mu}_{in}(x)+
{\frac{Z_\chi^{\frac{1}{2}}}{e_0{\ti v}}} \pa^\mu b_{in}(x)+
: F^{\mu}[\rho_{in}, U^\mu_{in}, \pa(\chi_{in}-b_{in})]:.
\ee
$ Z_\chi $, $ Z_\rho $ and $Z_3$ are the wave function renormalization constants.
As usual, the colon symbol denotes normal ordering.
The functionals  $F$
and $F^{\mu}$
have to be determined
within a specific choice for the Lagrangian. The $S$-matrix is given by $S\,=\,:
S[\rho_{in},
U^\mu_{in},
\pa(\chi_{in} - b_{in})]:$. It is important to notice that the dynamical mappings
(\ref{lp57a}) and (\ref{lp57b}) (and the one for the $S$-matrix)  are weak
equalities,
i.e. they are equalities among matrix elements computed in the Fock space for
the physical states. The free field equations are
\be\lab{lp24}
\pa^2 \chi_{in}(x)\,=\,0~,~~~
\pa^2 b_{in}(x)\,=\,0~, ~~~
(\pa^2 \, + \, m_\rho^2)\rho_{in}(x) \, =\,0~,
\ee
\be\lab{lp29}
( \pa^2 \, + \, {m_V}^{2}) U^\mu_{in}(x) \, =\, 0 \qquad, \qquad
\pa_{\mu} U^\mu_{in}(x) \, =\, 0~.
\ee
with ${m_V}^{2} =
\frac{Z_{3}}{Z_\chi}
(e_0{\ti v})^{2}$.

One can show that the local gauge transformations and the global
phase transformations (cf. (\ref{lp2}) and (\ref{lp1})) of the
Heisenberg fields are induced by the in-field transformations (see (\ref{lp57a})
and (\ref{lp57b})):
\be\lab{lp51a}
\chi_{in}(x)  \rar   \chi_{in}(x) \, + \, \frac{e_0 {\ti
v}}{Z_\chi^{\frac{1}{2}}} \la(x) ~,~~ ~
b_{in}(x)  \rar   b_{in}(x) \, + \, \frac{e_0 {\ti
v}}{Z_\chi^{\frac{1}{2}}} \la(x)~,
\ee
\be\lab{lp51c}
\rho_{in}(x)  \rar   \rho_{in}(x)~,~~  ~
U^\mu_{in}(x) \rar  U^\mu_{in}(x)  ~.
\ee
and by
\be\lab{lp53a}
\chi_{in}(x)  \rar  \chi_{in}(x) \, + \, \frac{{\ti
v}}{Z_\chi^{\frac{1}{2}}} \te f(x)   ~,
\ee
\be\lab{lp53b}
b_{in}(x) \rar b_{in}(x) \quad,
\rho_{in}(x) \rar \rho_{in}(x) \quad,
U^\mu_{in}(x) \rar U^\mu_{in}(x) ,
\ee
respectively. The square integrable function $f(x)$ is required to be
solution of $\pa^2 f(x) =0$. Note that (\ref{lp53a}) with $f(x)= 1$
(translation by a constant c-number) is not unitarily implementable.
$f(x)$ is introduced in order to make the generator of such a transformation well defined. The limit $f(x)\rar 1$ (i.e. the infinite volume limit) is to be performed at the end of the
computation. Notice that the in-field equations and the $S$ matrix
are invariant under the above in-field transformations (in the
limit $f \rar 1$ ).

The transformation (\ref{lp53a}) with $f(x)= 1$ describes the
{\it homogeneous boson condensation}. Once NG bosons are condensed in the
ground state, long range correlation is established which manifests as the
ordered pattern in the system ground state. This means that transitions between the system phases characterized by different ordered patterns in the ground state may be induced by the process of boson condensation. In other words, phase transitions (transitions among unitarily inequivalent Fock spaces) are induced by variations (gradients) of the NG boson
condensate.

The fact that the local and the global U(1) gauge transformations
are induced by the group $G'$ of transformations (\ref{lp51a}-\ref{lp53b})
is named {\it the dynamical rearrangement of symmetry} \cite{Um2, Um1}:
The dynamical rearrangement of symmetry
expresses the  consistency between the invariance of the Lagrangian under the symmetry transformations and the SBS condition $\lan 0|
\phi_H(x)|0\ran = {\ti v} \neq 0$.

Translations of boson fields (eqs. (\ref{lp51a})
and  (\ref{lp53a})) are thus obtained as a consequence of SBS.
As a general result one finds that $G'$
is the group contraction of the symmetry group of the dynamics (U(1)
in our present case; more generally, if the symmetry group is SU(n) or SO(n), the group contraction $G'$ is EU(n-1) or E(n), respectively) \cite{inonu}. Notice that $G'$ is the transformation group
relevant to the phase transitions process.

\bigskip

\bigskip

\bigskip

{\it Appendix B. Macroscopic field and currents generated by the boson condensation}

\vspace{0.5cm}

In the framework of the model in Appendix A, the Maxwell equations are
\be\lab{lp37}
- \pa^2 A_{H}^{\mu}(x) \, =\, j_{H}^{\mu}(x) \, -\,
\pa^{\mu} B(x) ~,
\ee
with
\be\lab{lp43}
B(x)= \frac{e_0 {\ti v}}{Z_\chi^{\frac{1}{2}}}[b_{in}(x) -
\chi_{in}(x)]  \, ~,~~~ \pa^2 B(x)\, =\,0   ~,
\ee
and with $j_{H}^{\mu}(x)= \de{\cal L}(x)/\de A_{H \mu} (x)$.
We may require that the current $j_{H}^{\mu}$ is the only source of
the gauge field $A_{H}^{\mu}$ in any observable process. This is obtained by
imposing the condition: $_p\lan b|\pa^{\mu} B(x)|a\ran_p\,
= \,0$, i.e.
\be\lab{lp45}
(- \pa^2) \,_p\lan b| {A^{0 \mu}}_{H}(x) |a \ran_p \, =
\,_p\lan b| j_{H}^{\mu}(x) |a\ran_p  ~,
\ee
where we use
$A^{0\mu}_{H}(x) \equiv A^{\mu}_{H}(x) -
{ e_0{\ti v}}:\pa^\mu b_{in}(x):$. $|a\ran_p $ and
$|b\ran_p $ denote two generic physical
states. The condition
$_p\lan b|\pa^{\mu} B(x)|a\ran_p\,
= \,0$ leads to the Gupta-Bleuler-like condition
\be\lab{lp49}
[\chi_{in}^{(-)}(x)  \, - \,  b_{in}^{(-)}(x)]|a\ran_p\, = \,0   ~,
\ee
with $\chi_{in}^{(-)}$ and $b_{in}^{(-)}$ the positive-frequency
parts of the corresponding fields. $\chi_{in}$ and
$b_{in}$ do not participate to any observable reaction. However,
we stress that
the NG bosons do not disappear from the theory: their
condensation in the vacuum can have observable effects.
Notice that {\it Eq.(\ref{lp45}) are the classical Maxwell equations}.

We now remark that the boson
transformation must be also compatible
with the physical state condition (\ref{lp49}).
Under the boson transformation $\chi_{in}(x)  \rar  \chi_{in}(x) +
\frac{{\ti
v}}{Z_\chi^{\frac{1}{2}}} f(x)$, $B$ changes as
\be\lab{vs9}
B(x) \rar B(x) - \frac{e_0 {\ti v}^2}{Z_\chi} f(x) ~.
\ee
Eq. (\ref{lp45}) is then violated. In order to
restore it, the shift in $B$ must be compensated by means of
the transformation on $U_{in}$:
\be\lab{vs10}
U^{\mu}_{in}(x) \rar U^{\mu}_{in}(x) +
{Z_{3}}^{-\frac{1}{2}} a^{\mu}(x) \qquad ,
\qquad \pa_\mu a^{\mu}(x)=0 ~,
\ee
with a convenient c-number function
$a^{\mu}(x)$.
The dynamical
maps of the various Heisenberg operators are not
affected by (\ref{vs10}) provided
\be\lab{Bvs20}
(\pa^2 + m_V^2) a_\mu(x) \, = \,\frac{m_V^2}{ e_0} \pa_\mu f(x)~.
\ee
{\it This is the Maxwell equation for
the vector potential $a_{\mu}$} \cite{MPUV75, MPU75}.
A similar result may be obtained in more complex models with non-Abelian symmetry groups \cite{Um1, MV90}.

\bigskip






\begin{thebibliography}{99}



\bibitem{Bunkov} Y.M. Bunkov and H. Godfrin (Eds.),
{\it Topological
defects and the non-equilibrium dynamics of
symmetry breaking phase transitions}, NATO
Science Series C 549, (Kluwer Acad. Publ. Dordrecht 2000).


\bibitem{kib} T.W.B. Kibble,
J. Phys.  {\bf A9}, 1387 (1976); Phys. Rep.
{\bf 67}, 183  (1980). \\
A. Vilenkin,
Phys. Rep.  {\bf 121}, 264  (1985).


\bibitem{kib2} T.W.B. Kibble,
in  {\it Topological
defects and the non-equilibrium dynamics of
symmetry breaking phase transitions}, eds.
Y.M. Bunkov and H. Godfrin, NATO
Science Series C 549, (Kluwer Acad. Publ. Dordrecht 2000), p. 7.

\bibitem{zurek1} W.H. Zurek,
Phys. Rep.  {\bf 276}, 177 (1997) and refs. therein quoted




\bibitem{volovik1} G.E. Volovik,
in {\em Topological
defects and the non-equilibrium dynamics of
symmetry breaking phase transitions}, eds.
Y.M. Bunkov and H. Godfrin, NATO
Science Series C 549, (Kluwer Acad. Publ. Dordrecht 2000), p. 353.

\bibitem{volovik2} G.E. Volovik,
in {\em Topological
defects and the non-equilibrium dynamics of
symmetry breaking phase transitions}, eds.
Y.M. Bunkov and H. Godfrin, NATO
Science Series C 549, (Kluwer Acad. Publ. Dordrecht 2000), p. 1.

\bibitem{zurek} W.H. Zurek, L.M.A.Bettencourt, J.Dziarmaga and
N.D.Antunes,
in {\em Topological
defects and the non-equilibrium dynamics of
symmetry breaking phase transitions}, eds.
Y.M. Bunkov and H. Godfrin, NATO
Science Series C 549, (Kluwer Acad. Publ. Dordrecht 2000), p. 77.

\bibitem{les} G. Vitiello,
in {\em Topological
defects and the non-equilibrium dynamics of
symmetry breaking phase transitions}, eds.
Y.M. Bunkov and H. Godfrin, NATO
Science Series C 549, (Kluwer Acad. Publ. Dordrecht 2000), p.171.

\bibitem{Um2} H. Umezawa,
{\em Advanced Field Theory:
Micro, Macro and Thermal Physics} (American Institute of Physics, 1993).

\bibitem{Um1} H. Umezawa, H. Matsumoto, and M. Tachiki,
{\em Thermo Field Dynamics and Condensed States},
(North-Holland Publ.Co., Amsterdam, 1982).


\bibitem{MV90} R. Manka and G. Vitiello,
Ann. Phys. (N.Y.) {\bf 199},  {61} (1990).

\bibitem{MPU75}
H. Matsumoto, N.J. Papastamatiou, and  H. Umezawa,
Nucl. Phys.  {\bf B97}, {90} (1975).


\bibitem{MPUV75}
H. Matsumoto, N.J. Papastamatiou, H. Umezawa and G. Vitiello,
{\it Nucl. Phys.} {\bf B97}, {61} (1975).


\bibitem{MPU741}
H. Matsumoto, N.J. Papastamatiou and H. Umezawa,
Nucl. Phys.  {\bf B68}, 236 (1974).

\bibitem{MPU742}
H. Matsumoto, N.J. Papastamatiou and H. Umezawa,
Nucl. Phys. {\bf B82}, 45 (1974).



\bibitem{itz} C. Itzykson and J.B. Zuber,
{\it Quantum Field Theory}
(MacGraw-Hill Book Co., N.Y. 1980)

\bibitem{hig} P. Higgs,
{Phys. Rev.}   {\bf 45}, {1156} (1960).

\bibitem{ki} T.W.B. Kibble,
Phys. Rev.  {\bf 155}, {1554} (1967).


\bibitem{inonu} C.~De Concini and G.~Vitiello,
Nucl. Phys. {\bf B116}, 141 (1976).




\end{thebibliography}
\end{document}